\documentclass[twocolumn]{aa} 
\usepackage{psfig,lscape}  
  
\def\kms{$\rm km\;s^{-1}$}  
\def\kmspc{$\rm km\;s^{-1}\;pc^{-1}$}
\def\dg{^\circ}  

\def\vs{$v_\star$}
\def\Vs{$V_\star$}
\def\vg{$v_g$}
\def\Vg{$V_g$}
\def\ss{$\sigma_\star$}
\def\sg{$\sigma_g$}
\def\htre{$h_{3}$}
\def\hqua{$h_{4}$}
 
\def\ha{H$\alpha$} 
\def\hi{H~{\small I}}

\def\oiii{[O~{\small III}]$\,\lambda5006.8$}  

\begin{document}  
  
  
\title{Kinematic properties of gas and stars in 20 disc galaxies  
\thanks{Based on observations carried out at the European Southern
Observatory, at the Multiple Mirror Telescope Observatory, at the
Observatorio del Roque de los Muchachos, at the Observatorio del Teide,
and at the Mount Graham International Observatory.}
$^{\bf,}$\thanks{Tables 5 and 6 are only available in electronic form
at the CDS via anonymous ftp to cdsarc.u-strasbg.fr (130.79.128.5) or
via http://cdsweb.u-strasbg.fr/Abstract.html.}}
  
\author{ 
     J.C.~Vega Beltr\'an  \inst{1,2}, 
     A.~Pizzella          \inst{3}, 
     E.M.~Corsini         \inst{3},  
     J.G.~Funes,~S.J.     \inst{4},  
     W.W.~Zeilinger       \inst{5}, 
     J.E.~Beckman         \inst{1},
     and F.~Bertola       \inst{6}}

\offprints{Juan Carlos Vega Beltr\'an}  
\mail{jvega@ll.iac.es}  
  
\institute{ 
Instituto Astrof\'\i sico de Canarias, Calle Via Lactea s/n, 
  E-38200 La Laguna, Spain \and 
Guest investigator of the UK Astronomy Data Centre \and 
Osservatorio Astrofisico di Asiago, Dipartimento di Astronomia, 
  Universit\`a di Padova, via dell'Osservatorio~8, I-36012 Asiago, Italy \and 
Vatican Observatory, University of Arizona, Tucson, AZ 85721, USA\and 
Institut f\"ur Astronomie, Universit\"at Wien, T\"urkenschanzstrasse 17, 
  A-1180 Wien, Austria\and 
Dipartimento di Astronomia, Universit\`a di Padova, 
  vicolo dell'Osservatorio~5, I-35122 Padova, Italy} 
  
\date{Version: \today }  
  
\titlerunning{Kinematic properties of gas and stars in 20 disc galaxies}  
\authorrunning{Vega Beltr\'an et al.}

\abstract{  
Ionized gas and stellar kinematical parameters have been measured
along the major axis of 20 nearby disc galaxies.  We discuss the
properties of each sample galaxy distinguishing between those
characterized by regular or peculiar kinematics. In early-type disc
galaxies ionized gas tends to rotate faster than stars and to have a
lower velocity dispersion (\Vg$\;>\;$\Vs\ and \sg$\;<\;$\ss), whereas
in late-type spirals gas and stars show almost the same rotation
velocities and velocity dispersions (\Vg$\;\simeq\;$\Vs\ and
\sg$\;\simeq\;$\ss).  Incorporating the early-type disc galaxies
studied by Bertola et al. (1995), Fisher  (1997) and Corsini et
al. (1999), we have compiled a sample of some 40 galaxies for which
the major-axis radial profiles of both the stellar and gaseous
components have been measured.  The value of \ss\ measured at $R_e/4$
turns out to be strongly correlated with the galaxy morphological
type, while \sg\ is not  and sometimes takes values above the range 
expected from thermal motions or small-scale turbulence.  
  }
\maketitle     
  
\section{Introduction}  
  
Our current understanding of galaxy formation has greatly benefited
from the results of $N-$body modelling of structure formation in the
early Universe, which predicts that small objects combine
gravitationally to produce the galaxies we see today: a process called
hierarchical-clustering-merging (hereafter HCM, cf. Kauffmann et
al. 1993). One of the tenets of the HCM paradigm is that galaxies are
constantly merging with one another.  In the case of elliptical and S0
galaxies, there is ample observational evidence that they are
continually subjected to mergers with smaller, neighbouring galaxies
(cf. Schweizer 1998).

If the HCM paradigm is universal, spirals are subject to the same
formation processes as E's and S0's. Often the fingerprints of such
second events reside in the stellar and/or gaseous kinematics of a
galaxy rather than in its morphology. This is particularly true if we
consider that the most evident `morphological tracers' of interactions
such as peculiar or spindle galaxies make up less than $5\%$ of all
objects in any one of the RC3 (de Vaucouleurs et al. 1991), UGC
(Nilson 1973) or ESO/Upssala (Lauberts 1982) galaxy catalogues.  It is
therefore crucial to obtain detailed kinematic parameters of both
stars and gas to unveil the relics of accretion or merging events
which have occurred in galaxy history.  A large fraction of spirals
exhibit kinematic disturbances ranging from mild to major, and can
generally be explained as the visible signs of tidal encounters
(Rubin, Waterman \& Kenney 1999). In recent years a number of
otherwise morphologically undisturbed spirals have been found which
host kinematically-decoupled components (KDC's), such as stellar KDC's
(Bertola et al. 1999; Sarzi et al. 2000), counter-rotating extended
stellar discs (Merrifield \& Kuijken 1994; Bertola et al. 1996; Jore
et al. 1996), counter-rotating or decoupled gaseous discs (Braun et
al. 1992; Rubin 1994; Rix et al. 1995; Ciri et al. 1995; Haynes et
al. 2000; Kannappan \& Fabricant 2001) and possibly counter-rotating
bulges (Prada et al. 1996; but see also Bottema 1999).

Studying the interplay between ionized gas and stellar kinematics
allows us to address other issues concerning the dynamical structure
of spirals. These include the origin of disc heating and the
presence of stellar or gaseous discs in galactic nuclei.
Gravitational scattering from giant molecular clouds and spiral
density waves are the prime candidates to explain the finite thickness
of stellar discs.  It is expected that the dominant heating mechanism
varies along the Hubble sequence but up to now only two external
galaxies have been studied in detail (Gerssen, 
Kuijken \& Merrifield 1996, 2000).
The presence in the nuclei of S0's and spirals of small stellar
(Emsellem et al. 1996; Kormendy et al. 1996a,b; van den Bosch, Jaffe
\& van der Marel 1998; Scorza \& van den Bosch 1998; van den Bosch \&
Emsellem 1998) and/or gaseous discs (Rubin, Kenney, \& Young 1997;
Bertola et al. 1998; Funes 2000) is usually connected to the presence
of a central mass concentration. It also appears that the central
black-hole mass is very strongly correlated with the stellar velocity
dispersion of the host galaxy bulge as recently found by
different authors (Ferrarese \& Merritt 2000; Gebhardt et
al. 2000). This relation is however based on samples which are
affected by different biases and therefore new black-hole masses as
well as stellar velocity dispersion measurements are needed.

Finally, the comparison of mass distributions derived from stellar and
gaseous kinematics has shown that the ionized gas velocity may not
trace the circular speed in the central regions of S0's (Fillmore,
Boroson \& Dressler 1986; Bertola et al. 1995; Cinzano et al. 1999)
and bulge-dominated spirals (Corsini et al. 1999; Pignatelli et
al. 2001). The possible difference between the gas rotational velocity
and the gravitational equilibrium circular velocity poses questions
about the reliability of mass distributions derived from the direct
decomposition of ionized gas rotation cur\-ves into the bulge, disc
and dark halo contribution (see Kent 1988 for a discussion). This
phenomenon has been explained in terms of pressure-supported ionized
gas, gas motions which are not confined to the galaxy equatorial plane
and drag forces but its cause is still unclear due to the limited
statistics and requires further investigation.

All these issues will benefit greatly from a survey devoted to the
comparative measurements of ionized gas and stellar kinematics.  With
this aim we obtained long-slit spectroscopy of a sample of 20 disc
galaxies,  mostly spirals.  We measured the velocity,
velocity dispersion, \htre\ and \hqua\ radial profiles of the stellar
component and velocity and velocity dispersion radial profiles of the
ionized gas along their major axes. In Pignatelli et al. (2001) we
present the mass modelling of three galaxies of the sample, the Sa NGC
772 and the Sb's NGC 3898 and NGC 7782.

This paper is organized as follows.  An overview of the properties
of the sample galaxies as well as the spectroscopic observations and
their data analysis are presented in Sect. \ref{sec:observations}.  The
resulting stellar and gaseous kinematic parameters are given in
Sect. \ref{sec:results}. Conclusions are discussed in
Sect. \ref{sec:conclusions}. In the appendix a comparison
with published kinematic measurements of the sample galaxies is
performed.

\begin{table*}[t]  
\caption[]{Basic properties of the sample galaxies}  
\begin{center}  
\begin{footnotesize}  
\begin{tabular}{lllrrcrrrrr}  
\hline  
\noalign{\smallskip}  
\multicolumn{1}{c}{object} &  
\multicolumn{2}{c}{type} &  
\multicolumn{1}{c}{$B_T$} &  
\multicolumn{1}{c}{P.A.} &  
\multicolumn{1}{c}{$i$} &  
\multicolumn{1}{c}{$V_{\odot}$} &  
\multicolumn{1}{c}{$D$} &  
\multicolumn{1}{c}{scale} &  
\multicolumn{1}{c}{$R_{25}$} & 
\multicolumn{1}{c}{$M_{B_T}^0$} \\   
\multicolumn{1}{c}{[name]} &  
\multicolumn{1}{c}{[RSA]} &  
\multicolumn{1}{c}{[RC3]} &  
\multicolumn{1}{c}{[mag]} &  
\multicolumn{1}{c}{[\degr]} &  
\multicolumn{1}{c}{[\degr]} &  
\multicolumn{1}{c}{[\kms]} &  
\multicolumn{1}{c}{[Mpc]} &  
\multicolumn{1}{c}{[pc$/''$]} &  
\multicolumn{1}{c}{[$'$]} & 
\multicolumn{1}{c}{[mag]} \\  
\multicolumn{1}{c}{(1)} &  
\multicolumn{1}{c}{(2)} &  
\multicolumn{1}{c}{(3)} &  
\multicolumn{1}{c}{(4)} &  
\multicolumn{1}{c}{(5)} &  
\multicolumn{1}{c}{(6)} &  
\multicolumn{1}{c}{(7)} &  
\multicolumn{1}{c}{(8)} &  
\multicolumn{1}{c}{(9)} &  
\multicolumn{1}{c}{(10)} &  
\multicolumn{1}{c}{(11)} \\
\noalign{\smallskip}  
\hline  
\noalign{\smallskip}  
NGC~224  & Sb     &.SAS3..      &  4.36 &  55 & 72 &$-290$&  0.7 &   3.4 &95.3 &$-20.87$\\ 
NGC~470  & Sbc(s) &.SAT3..      & 12.53 & 155 & 52 & 2370 & 33.8 & 163.9 & 1.4 &$-20.66$\\  
NGC~772  & Sb(rs) &.SAS3..      & 11.09 & 130 & 54 & 2470 & 35.6 & 172.7 & 3.6 &$-22.21$\\
NGC~949  & Sc(s)  &.SAT3$\ast$\$& 12.40 & 145 & 58 &  620 & 11.4 &  55.2 & 1.2 &$-18.50$\\     
NGC~980  & ...    &.L.....      & 13.20 & 110 & 58 & 5765 & 80.1 & 388.2 & 0.8 &$-22.95$\\  
NGC~1160 & ...    &.S..6$\ast$. & 13.50 &  50 & 62 & 2510 & 36.6 & 177.4 & 1.0 &$-21.01$\\
NGC~2541 & Sc(s)  &.SAS6..      & 12.26 & 165 & 61 &  565 &  8.7 &  42.2 & 3.2 &$-18.13$\\
NGC~2683 & Sb     &.SAT3..      & 10.64 &  44 & 78 &  460 &  5.3 &  25.6 & 4.7 &$-18.99$\\  
NGC~2841 & Sb     &.SAR3$\ast$. & 10.09 & 147 & 65 &  640 &  9.6 &  46.4 & 4.1 &$-20.33$\\  
NGC~3031 & Sb(r)  &.SAS2..      &  7.89 & 157 & 59 & $-30$&  1.5 &   7.2 &13.5 &$-18.46$\\ 
NGC 3200 & Sb(r)  &.SXT5$\ast$. & 12.83 & 169 & 73 & 3550 & 43.9 & 213.1 & 2.1 &$-21.53$\\  
NGC~3368 & Sab(s) &.SXT2..      & 10.11 &   5 & 47 &  860 &  9.7 &  47.1 & 3.8 &$-20.14$\\  
NGC~3705 & Sab(r) &.SXR2..      & 11.86 & 122 & 66 & 1000 & 11.4 &  55.2 & 2.4 &$-19.03$\\  
NGC~3810 & Sc(s)  &.SAT5..      & 11.35 &  15 & 45 & 1000 & 11.9 &  56.0 & 2.1 &$-19.36$\\  
NGC~3898 & Sa     &.SAS2..      & 11.60 & 107 & 54 & 1185 & 17.1 &  82.9 & 2.2 &$-19.85$\\  
NGC~4419 & SBab:  &.SBS1./      & 12.08 & 133 & 71 &$-200$& 17.0 &  82.4 & 1.7 &$-19.55$\\  
NGC~5064 & Sa     &PSA.2$\ast$. & 13.04 &  38 & 64 & 2980 & 36.0 & 174.4 & 1.2 &$-21.11$\\  
NGC~5854 & Sa     &.LBS$+$./    & 12.71 &  55 & 76 & 1630 & 21.8 & 100.7 & 1.4 &$-18.90$\\ 
NGC~7331 & Sb(rs) &.SAS3..      & 10.35 & 171 & 70 &  820 & 14.7 &  72.0 & 5.2 &$-20.48$\\  
NGC~7782 & Sb(s)  &.SAS3..      & 13.08 & 175 & 58 & 5430 & 75.3 & 364.9 & 1.2 &$-21.95$\\     
\noalign{\smallskip}  
\hline  
\noalign{\smallskip}  
\noalign{\smallskip}  
\noalign{\smallskip}  
\end{tabular}  
\begin{minipage}{18cm}  
NOTES -- Col.(2): morphological classification from RSA.
Col.(3): morphological classification from RC3.
Col.(4): total observed blue magnitude from RC3 except for 
         NGC~980 and NGC 5064 (LEDA). 
Col.(5): major-axis position angle taken from RC3. 
Col.(6): inclination derived as 
         $\cos^{2}{i}\,=\,(q^2-q_0^2)/(1-q_0^2)$. The observed axial ratio $q$ 
         is taken from RC3 and the intrinsic flattening $q_0=0.11$ has been 
         assumed following Guthrie (1992). 
Col.(7): heliocentric velocity of the galaxy derived at centre of 
         symmetry of the rotation curve of the gas. $\Delta V_\odot = 10$
         \kms .
Col.(8): distance obtained as $V_0/H_0$ with $H_0=75$ \kms\ 
         Mpc$^{-1}$ and $V_0$ the systemic velocity derived from $V_\odot$ 
         corrected for the motion of the Sun with respect to the Local Group 
         as in the RSA.
         For NGC~224 and NGC~4419 we assume distances of 
         0.7 Mpc (Binney \& Merrifield 1999) 
         and 17 Mpc (Freedman et al. 1994), respectively.  
Col.(10): radius of the 25 $B-$mag arcsec$^{-2}$ isophote derived as  
          $R_{25} = D_{25}/2$ with $D_{25}$ from RC3. 
Col.(11): absolute total blue magnitude corrected for
          inclination and extinction from RC3.
\end{minipage}  
\end{footnotesize}    
\end{center}  
\label{tab:sample}  
\end{table*}

\section{Sample selection, observations and data reduction} 
\label{sec:observations}

\begin{figure} 
\centerline{{\psfig{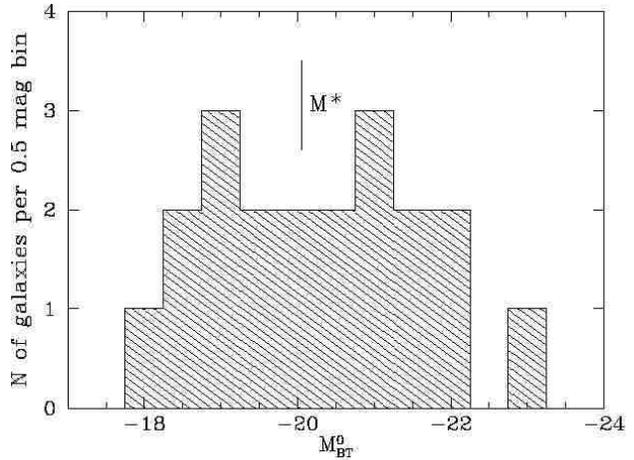}}} 
\caption{Absolute magnitude distribution for the sample galaxies.
A line marks $M_{B_T}^0 = -20.05$, which corresponds to  
M$^\ast$ for spiral galaxies as derived by Marzke et al. (1998)
and assuming H$_0 = 75$ \kms\ Mpc$^{-1}$.} 
\label{fig:histogram} 
\end{figure}

All the observed galaxies are bright ($B_T\leq13.5$) and nearby
objects ($V_\odot < 5800$ \kms) with an intermediate-to-high
inclination ($45\dg \leq i \leq 80\dg$) and their Hubble morphological
type ranges from S0 to Sc. An overview of their basic properties is
given in Tab.~\ref{tab:sample}.  Fig. \ref{fig:histogram} shows the
absolute magnitude distribution for the galaxies of our sample. It
nicely brackets the M$^\ast$ value for spiral galaxies taken from
Marzke et al. (1998) assuming H$_0 = 75$ \kms\ Mpc$^{-1}$.

\subsection{Spectroscopic observations} 
\label{sec:spectroscopy} 
 
The long-slit spectroscopic observations of our sample ga\-laxies were
carried out at the 4.5-m Multi Mirror Telescope (MMT) in Arizona (USA), 
at the ESO 1.52-m Spectroscopic Telescope at La Silla (Chile), 
and at the 2.5-m Isaac Newton Telescope (INT) on La Palma
(Spain). The instrumental setup of each observing run is
summarized in Tab. \ref{tab:setup}.

\begin{table*}[ht!] 
\caption{Instrumental setup of spectroscopic observations} 
\begin{tabular}{lccccc} 
\hline 
\noalign{\smallskip} 
\multicolumn{1}{c}{Parameter} &
\multicolumn{2}{c}{MMT} & 
\multicolumn{1}{c}{ESO 1.52-m} & 
\multicolumn{1}{c}{INT} \\ 
\noalign{\smallskip} 
\hline 
\noalign{\smallskip} 
Date                  & 21-23 Oct 1990 & 17-18 Dec 90             & 30 Apr - 02 May 1992 & 20-21 Mar 1996\\ 
Spectrograph          & \multicolumn{2}{c}{Red Channel}           & B\&C                 & IDS\\           
Grating ($\rm grooves\;mm^{-1}$) & \multicolumn{2}{c}{1200}       & 1200 (ESO No. 26)    & 1800\\           
Detector              & \multicolumn{2}{c}{Loral 12$\times$8mmt}  & FA2048L (ESO No. 24) & TK1024A\\
Pixel size ($\rm \mu m^{2}$) & \multicolumn{2}{c}{$15\times15$}   & $15\times15$         & $24\times24$\\
Pixel binning         &  \multicolumn{2}{c}{$1\times1$}           & $1\times1$           & $1\times1$\\  
Scale ($\rm ''\;pixel^{-1}$)  & \multicolumn{2}{c}{0.30}          &  0.81                & 0.33\\            
Reciprocal dispersion ($\rm \AA\;pixel^{-1}$) & \multicolumn{2}{c}{0.82} & 0.98          & 0.24\\ 
Slit width ($''$)   & \multicolumn{2}{c}{1.25}                    & 2.1                  & 1.9\\
Slit length ($'$)   & \multicolumn{2}{c}{3.0}                     & 4.2                  & 4.0\\
Spectral range (\AA)     & \multicolumn{2}{c}{4850--5500}         & 4900--6900           & 6650--6890\\
Comparison lamp      & \multicolumn{2}{c}{He--Ne--Ar--Fe}         & He--Ar               & Cu--Ar\\  
Instrumental FWHM (\AA) & $2.24\pm0.26$  & $2.57\pm0.11$          & $2.34\pm0.09$        & $0.869\pm0.040$\\ 
Instrumental $\sigma$ (\kms) & 57 & 65                            & 45                   & 17 \\ 
Seeing FWHM ($''$)           & \multicolumn{2}{c}{1.2--1.5}       & 1.0--1.5             & 1.0--1.8\\ 
\noalign{\smallskip} 
\hline 
\noalign{\smallskip} 
\noalign{\smallskip} 
\noalign{\smallskip} 
\end{tabular} 
\begin{minipage}{18cm} 
NOTES -- The instrumental $\sigma$ was measured at \oiii\ 
for the MMT spectra and at \ha\ for the ESO 1.52-m and INT spectra.
\end{minipage}
\label{tab:setup}  
\end{table*}

At the beginning of each exposure, the slit was centred on the galaxy
nucleus using the guiding TV camera and aligned along the galaxy major
axis. The details of the slit position and spectra exposure times are
given in Tab. \ref{tab:log_spectroscopy_galaxies}. In all the observing runs
comparison lamp exposures were obtained before and after each object
integration to allow an accurate wavelength calibration.  Quartz-lamp
and twilight-sky flat fields were used to map pixel-to-pixel
sensitivity variations and large-scale illumination patterns. 
At the MMT and ESO 1.52-m telescopes a number of late-G and early-K
stars were observed with the same
set up to serve as templates in measuring the stellar kinematics
(see Tab. ~\ref{tab:log_spectroscopy_stars}).
The seeing range during the different 
spectroscopic runs is given in Tab. \ref{tab:setup}.

\begin{table}
\caption[]{Log of spectroscopic observations of the galaxies}  
\begin{flushleft}  
\begin{tabular}{lllcr}  
\hline  
\noalign{\smallskip}   
\multicolumn{1}{c}{Object} &   
\multicolumn{1}{c}{Date}&  
\multicolumn{1}{c}{Telescope}&   
\multicolumn{1}{c}{$t_{\it exp}$} &   
\multicolumn{1}{c}{P.A.}\\  
\multicolumn{1}{c}{} &   
\multicolumn{1}{c}{} &    
\multicolumn{1}{c}{} &   
\multicolumn{1}{c}{[s]} &  
\multicolumn{1}{c}{[$\dg$]} \\  
\noalign{\smallskip}   
\hline  
\noalign{\smallskip}   
NGC~224  & 18 Dec 90 & MMT       & 2$\times$3600 &  55 \\   
NGC~470  & 22 Oct 90 & MMT       & 3600          & 155 \\   
NGC~772  & 22 Oct 90 & MMT       & 3600          & 130 \\   
NGC~949  & 21 Oct 90 & MMT       & 3600          & 145 \\   
NGC~980  & 22 Oct 90 & MMT       & 3600          & 110 \\  
NGC~1160 & 21 Oct 90 & MMT       & 3600          &  50 \\    
NGC~2541 & 21 Oct 90 & MMT       & 3600          & 165 \\  
NGC~2683 & 18 Dec 90 & MMT       & 3600          &  44 \\  
NGC~2841 & 22 Oct 90 & MMT       & 3600          & 147 \\  
NGC~3031 & 17 Dec 90 & MMT       & 3600          & 157 \\  
NGC~3200 & 02 May 92 & ESO 1.52-m& 3600          &  79 \\  
NGC~3368 & 17 Dec 90 & MMT       & 3600          &   5 \\  
NGC~3705 & 17 Dec 90 & MMT       & 3600          & 122 \\  
NGC~3810 & 18 Dec 90 & MMT       & 3600          &  15 \\  
NGC~3898 & 18 Dec 90 & MMT       & 3600          & 107 \\  
         & 19 Mar 96 & INT       & 3$\times$3600 & 107 \\  
NGC~4419 & 20 Mar 96 & INT       & 2$\times$3300 & 133 \\  
         & 02 May 92 & ESO 1.52-m& 3600          & 133 \\  
NGC~5064 & 30 Apr 92 & ESO 1.52-m& 3600          & 138 \\  
NGC~5854 & 30 Apr 92 & ESO 1.52-m& 3600          & 145 \\  
         & 30 Apr 92 & ESO 1.52-m& 3600          &  25 \\  
         & 30 Apr 92 & ESO 1.52-m& 3600          &  55 \\  
NGC~7331 & 22 Oct 90 & MMT       & 3600          & 171 \\  
NGC~7782 & 22 Oct 90 & MMT       & 3600          &  30 \\  
\noalign{\smallskip}   
\hline  
\end{tabular}
\label{tab:log_spectroscopy_galaxies}  
\end{flushleft}  
\end{table}

\begin{table}
\caption[]{Log of spectroscopic observations of the template stars}  
\begin{flushleft}  
\begin{tabular}{lllcl}  
\hline  
\noalign{\smallskip}   
\multicolumn{1}{c}{Object}    &   
\multicolumn{1}{c}{Date}      &  
\multicolumn{1}{c}{Telescope} & 
\multicolumn{1}{c}{$t_{\it exp}$} &   
\multicolumn{1}{c}{Type}      \\  
\multicolumn{1}{c}{}  &   
\multicolumn{1}{c}{}  &    
\multicolumn{1}{c}{}  &   
\multicolumn{1}{c}{[s]}  &   
\multicolumn{1}{c}{[BSC]} \\  
\noalign{\smallskip} 
\hline  
\noalign{\smallskip}   
HR 2649 & 21 Oct 90 & MMT  & 180 & K3III \\  
HR 7778 & 23 Oct 90 & MMT  & 200 & G8III \\  
HR 7854 & 23 Oct 90 & MMT  & 200 & K0III \\  
\noalign{\smallskip}                       
HR 941  & 18 Dec 90 & MMT  &  99 & K0III \\  
HR 3360 & 17 Dec 90 & MMT  & 195 & K2III \\  
HR 3905 & 17 Dec 90 & MMT  & 126 & K2III \\  
HR 4246 & 18 Dec 90 & MMT  & 100 & K3III \\  
HR 4924 & 17 Dec 90 & MMT  & 139 & G9III \\  
HR 8694 & 18 Dec 90 & MMT  &  89 & K0III \\  
\noalign{\smallskip}                       
HR 3431 & 30 Apr 92 & ESO 1.52-m & 5$\times$20 & K4III \\   
HR 5601 & 30 Apr 92 & ESO 1.52-m & 5$\times$15 & K0.5III \\   
HR 6318 & 30 Apr 92 & ESO 1.52-m & 5$\times$20 & K4III \\   
HR 7595 & 30 Apr 92 & ESO 1.52-m & 5$\times$20 & K0III \\     
\noalign{\smallskip}   
\hline  
\noalign{\smallskip}  
\noalign{\smallskip}  
\noalign{\smallskip}  
\end{tabular}  
\begin{minipage}{9cm}  
NOTE -- The spectral class of the template star is taken from The  
        Bright Star Catalogue (Hoffleit \& Jaschek 1982).   
\end{minipage}  
\label{tab:log_spectroscopy_stars}  
\end{flushleft}  
\end{table}

\subsection{Routine data reduction} 
\label{sec:data_reduction} 
 
The spectra were bias subtracted, flat-field corrected, cleaned for
cosmic rays and wavelength calibrated using standard
MIDAS\footnote{MIDAS is developed and maintained by the European
Southern Observatory.} routines.  Cosmic rays were identified by
comparing the counts in each pixel with the local mean and standard
deviation (as obtained from the Poisson statistics of the photons
knowing the gain and readout noise of the detector), and then
corrected by interpolating a suitable value.

The instrumental resolution was derived as the mean of the Gaussian
FWHM's measured for a dozen unblended arc-lamp lines distributed over
the whole spectral range of a wavelength-calibrated comparison
spectrum.  The mean FWHM of the arc-lamp lines as well as the
corresponding instrumental velocity dispersion are given in Tab.
\ref{tab:setup}.  Finally, the individual spectra of the same object
were aligned and coadded using their stellar-continuum centres as
reference. For each spectrum the centre of the galaxy was defined by
the centre of a Gaussian fit to the radial profile of the stellar
continuum.  The contribution of the sky was determined from the edges
of the resulting spectrum and then subtracted.

\subsection{Measuring stellar and ionized gas kinematics} 
\label{sec:measuring_kinematics} 

The stellar kinematic parameters were measured from the absorption
lines present on each spectrum using the Fourier Correlation Quotient
Method (Bender 1990) as applied by Bender, Saglia \& Gerhard (1994).
The spectra of the stars G8III HR 7778, K2III HR 6415 and K4III HR
6318 provided the best match to the galaxy spectra obtained in October
1990, December 1990 and May 1992, respectively.  They were used as
templates to measure the stellar kinematic parameters of the sample
galaxies in the three runs.  For each spectrum we measured the radial
profiles of the heliocentric stellar velocity (\vs), velocity
dispersion (\ss), and the Gauss-Hermite coefficients \htre\ and \hqua,
in the case of sufficiently high S/N.
The stellar kinematics of all the sample galaxies are tabulated in
Tab. 5.  The table provides the galaxy name, the position angle of the
slit in degrees, the radial distance from the galaxy centre in arcsec,
the observed heliocentric velocity and the velocity dispersion in
\kms, and the Gauss-Hermite coefficients $h_3$ and $h_4$.

The ionized gas kinematic parameters were derived by measuring the
position and the width of \oiii\ emission line in the MMT spectra and
the \ha\ emission line in the ESO 1.52-m and INT spectra.  The
position, the FWHM and the uncalibrated flux of the emission lines
were individually determined by fitting interactively a single
Gaussian to each emission line, and a polynomial to its surrounding
continuum using the MIDAS package ALICE.  The wavelength of the
Gaussian peak was converted to velocity via the optical convention
$v=cz$, and then the standard heliocentric correction was applied to
obtain the ionized gas heliocentric velocity (\vg). The Gaussian FWHM
was corrected for the instrumental FWHM, and then converted to
velocity dispersion (\sg). At some radii where the intensity of the
emission lines was low, we averaged adjacent spectral rows to improve
the signal-to-noise ratio of the lines.
The ionized-gas kinematic parameters of all the sample galaxies are
tabulated in Tab. 6.  The table provides the galaxy name, the position
angle of the slit in degrees, the radial distance from the galaxy
centre in arcsec, the observed heliocentric velocity and the velocity
dispersion in \kms, and the relevant emission line.
For each galaxy we derive the heliocentric system velocity as the velocity of
the centre of symmetry of the rotation curve of the gas.

\section{Results}  
\label{sec:results}

\begin{figure*}
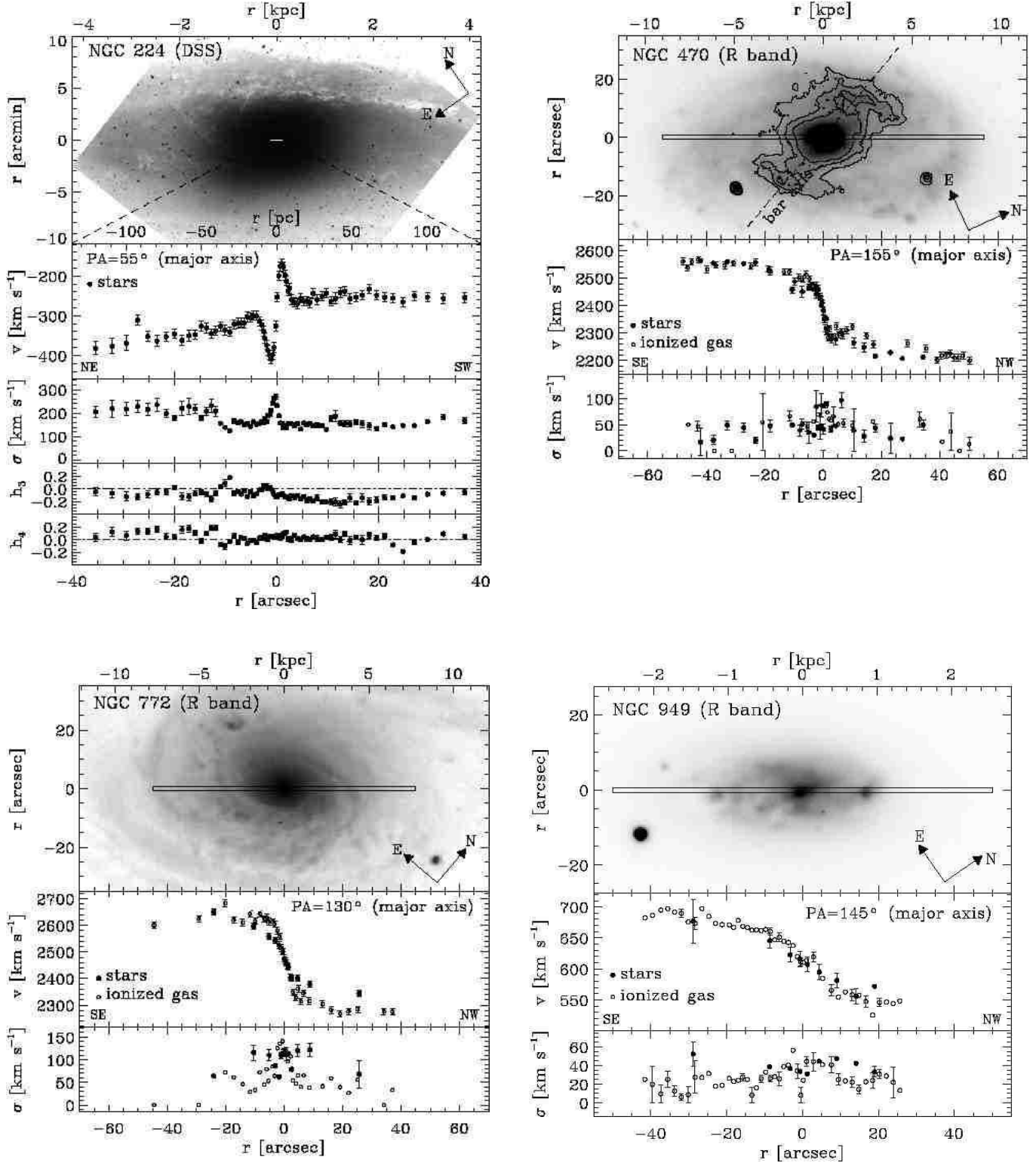

\begin{minipage}[t]{8.5cm} 
\vspace{0pt}  
\psfig{figure=MS10597f2a.eps1,width=8.5cm}
\end{minipage}
\hspace*{0.5cm}
\begin{minipage}[t]{8.5cm}
\vspace{0pt}
\psfig{figure=MS10597f2b.eps1,width=8.5cm}
\end{minipage}
\vspace*{0.cm}
\begin{minipage}[t]{8.5cm}   
\vspace{20pt}
\psfig{figure=MS10597f2c.eps1,width=8.5cm}
\end{minipage}
\hspace*{0.5cm}
\begin{minipage}[t]{8.5cm}
\vspace{20pt}
\psfig{figure=MS10597f2d.eps1,width=8.5cm}
\end{minipage}
\caption{Stellar and ionized-gas kinematics of the sample galaxies. For NGC~470 some isophotes are traced to enlight the presence of the bar,
whose major axis is drawn with a dashed line}
\label{fig:kinematics}  
\end{figure*}

The resulting stellar and ionized-gas kinematics of all our
sample galaxies are shown in Fig. \ref{fig:kinematics}.  Their
relevant kinematic properties such as velocity gradients and velocity
dispersions for both the stellar and gaseous components at different
radii are given in Tab. \ref{tab:results}.

For each object the plot of the stellar and gaseous kinematics is
organized as it follows:

\begin{enumerate}  
\item We display in the upper panel an image of the galaxy.
  We obtained the images for most of the sample galaxies in February
  and November 1997 at the 1.8-m Vatican Advanced Technology (VATT)
  in Arizona (USA), in December 1997 at the 0.8-m IAC80 telescope in
  Tenerife (Spain), and in March 1998 at the ESO 3.6-m telescope in La
  Silla (Chile).  For the remaining galaxies we used an image taken
  either from the INT archive or the Digital Sky Survey (see
  Table \ref{tab:images}). All images (except for the DSS ones) were
  bias-subtracted, flat-field corrected and cleaned for cosmic rays.
  The galaxy frames have been rotated and the slit position has been
  plotted to perform a better comparison between morphological and
  kinematic properties. The slit width and length correspond to those adopted
  in obtaining the spectra. 

\setcounter{table}{6} 
\begin{table}
\caption[]{Source of the images of sample galaxies}  
\begin{flushleft}  
\begin{tabular}{llc}  
\hline  
\noalign{\smallskip}   
\multicolumn{1}{c}{Object}    &   
\multicolumn{1}{c}{Source}    &    
\multicolumn{1}{c}{Band}     \\
\noalign{\smallskip}   
\hline  
\noalign{\smallskip}   
NGC~224  & DSS         &      \\ 
NGC~470  & IAC80       & $R_J$\\
NGC~772  & VATT        & $R_C$\\
NGC~949  & VATT        & $R_C$\\
NGC~980  & IAC80       & $R_J$\\  
NGC~1160 & VATT        & $R_C$\\
NGC~2541 & IAC80       & $R_J$\\   
NGC~2683 & INT Archive & $R_C$\\
NGC~2841 & IAC80       & $R_J$\\
NGC~3031 & DSS         &      \\
NGC 3200 & DSS         &      \\
NGC~3368 & INT Archive & $R_C$\\
NGC~3705 & DSS         &      \\
NGC~3810 & DSS         &      \\
NGC~3898 & VATT        & $R_C$\\
NGC~4419 & VATT        & $R_C$\\
NGC~5064 & ESO 3.6-m   & $R_C$\\
NGC~5854 & IAC80       & $R_J$\\
NGC~7331 & INT Archive & $R_C$\\
NGC~7782 & VATT        & $R_C$\\
\noalign{\smallskip}   
\hline  
\end{tabular}  
\label{tab:images}  
\end{flushleft}  
\end{table}  

\item In the second panel we plot the velocity curves of the stellar
  (filled circles) and gaseous (open circles) components.  The 
  velocities are as observed without any inclination correction. Error
  bars are not plotted when smaller than symbols. 
  The position angle of the slit is specified.

\item In the third panel we plot in radial profiles the velocity
  dispersion of the stellar (filled circles) and gaseous 
  (open circles) components. Error bars are not plotted when smaller than symbols.
 
\item In the lower panels we plot the \htre\ and \hqua\
  radial profiles, respectively. \htre\ and \hqua\ values are not given
  where the $S/N$ ratio of the spectra was too low to allow a 
  reliable measurement. 
\end{enumerate}

A detailed study of the ionized gas and stellar kinematics of NGC 772,
NGC 3898 and NGC 7782 is given in Pignatelli et al. (2001). They
combined kinematical data with $V-$band surface photometry in order to
derive the mass distribution via a self-consistent Jeans model.
Kinematic data for these galaxies are presented here for sake of
completeness.

The kinematic parameters of the nearby galaxies NGC 224 and NGC 3031
have been measured by several authors. We decided to observe these
galaxies to perform a consistency check of our measurements with the
data available in the literature.

\addtocounter{figure}{-1}

\begin{figure*}
\begin{minipage}[t]{8.5cm} 
\vspace{0pt}  
\psfig{figure=MS10597f2e.eps1,width=8.5cm}
\end{minipage}
\hspace*{0.5cm}
\begin{minipage}[t]{8.5cm}
\vspace{0pt}
\psfig{figure=MS10597f2f.eps1,width=8.5cm}
\end{minipage}
\vspace*{0.cm}
\begin{minipage}[t]{8.5cm}   
\vspace{20pt}
\psfig{figure=MS10597f2g.eps1,width=8.5cm}
\end{minipage}
\hspace*{0.5cm}
\begin{minipage}[t]{8.5cm}
\vspace{20pt}
\psfig{figure=MS10597f2h.eps1,width=8.5cm}
\end{minipage}
\caption{(continue)}
\end{figure*} 

\addtocounter{figure}{-1}

\begin{figure*}
\begin{minipage}[t]{8.5cm} 
\vspace{0pt}  
\psfig{figure=MS10597f2i.eps1,width=8.5cm}
\end{minipage}
\hspace*{0.5cm}
\begin{minipage}[t]{8.5cm}
\vspace{0pt}
\psfig{figure=MS10597f2j.eps1,width=8.5cm}
\end{minipage}
\vspace*{0.cm}
\begin{minipage}[t]{8.5cm}   
\vspace{20pt}
\psfig{figure=MS10597f2k.eps1,width=8.5cm}
\end{minipage}
\hspace*{0.5cm}
\begin{minipage}[t]{8.5cm}
\vspace{20pt}
\psfig{figure=MS10597f2l.eps1,width=8.5cm}
\end{minipage}
\caption{(continue)}
\end{figure*} 

\addtocounter{figure}{-1}

\begin{figure*}
\begin{minipage}[t]{8.5cm} 
\vspace{0pt}  
\psfig{figure=MS10597f2m.eps1,width=8.5cm}
\end{minipage}
\hspace*{0.5cm}
\begin{minipage}[t]{8.5cm}
\vspace{0pt}
\psfig{figure=MS10597f2n.eps1,width=8.5cm}
\end{minipage}
\vspace*{0.cm}
\begin{minipage}[t]{8.5cm}   
\vspace{20pt}
\psfig{figure=MS10597f2o.eps1,width=8.5cm}
\end{minipage}
\hspace*{0.5cm}
\begin{minipage}[t]{8.5cm}
\vspace{20pt}
\psfig{figure=MS10597f2p.eps1,width=8.5cm}
\end{minipage}
\caption{(continue)}
\end{figure*} 

\addtocounter{figure}{-1}

\begin{figure*}
\begin{minipage}[t]{8.5cm} 
\vspace{0pt}  
\psfig{figure=MS10597f2q.eps1,width=8.5cm}
\end{minipage}
\hspace*{0.5cm}
\begin{minipage}[t]{8.5cm}
\vspace{0pt}
\psfig{figure=MS10597f2r.eps1,width=8.5cm}
\end{minipage}
\vspace*{0.cm}
\begin{minipage}[t]{8.5cm}   
\vspace{20pt}
\psfig{figure=MS10597f2s.eps1,width=8.5cm}
\end{minipage}
\hspace*{0.5cm}
\begin{minipage}[t]{8.5cm}
\vspace{20pt}
\psfig{figure=MS10597f2t.eps1,width=8.5cm}
\end{minipage}
\caption{(continue)}
\end{figure*} 

\addtocounter{figure}{-1}

\begin{figure*}
\begin{minipage}[t]{8.5cm} 
\vspace{0pt}  
\psfig{figure=MS10597f2u.eps1,width=8.5cm}
\end{minipage}
\hspace*{0.5cm}
\begin{minipage}[t]{8.5cm}
\vspace{0pt}
\psfig{figure=MS10597f2v.eps1,width=8.5cm}
\end{minipage}
\caption{(continue)}
\end{figure*} 
 
\section{Discussion and conclusions}  
\label{sec:conclusions}

Although only a complete dynamical model can address the question of
the mass distribution of a galaxy, it is possible to derive some hints
about its structure directly from the analysis of the interplay
between the kinematics of its gas and stars.  In our sample we can
identify two classes of galaxies according to their kinematics,
assuming that gas and stars are coplanars:

\medskip 

\noindent
{\it (i)} Galaxies in which ionized gas rotates faster than stars and
has a lower velocity dispersion than the stars (i.e., \Vg$\;>\;$\Vs\
and \sg$\;<\;$\ss): NGC 772, NGC 3200, NGC 3898, NGC 4419, NGC 5064
and NGC 7782. All these galaxies are classified as
early-to-intermediate type spirals, except for the Sc NGC 3200.  The
different kinematic behaviour of the gaseous and stellar components
can be easily explained by a model where the gas is confined in the
disc and supported by rotation while the stars mostly belong to the
bulge and are supported by random motions (i.e. dynamical pressure).
In the case of NGC 772 and NGC 7782, this simple hypothesis is
confirmed by the self-consistent Jeans models of Pignatelli et
al. (2001).  In these galaxies the ionized gas is tracing the
gravitational equilibrium circular speed. This is not true in the
innermost region ($\pm0.7$ kpc) of NGC 3898, where the ionized gas is
rotating more slowly than the circular velocity predicted from
dynamical modelling, unveiling a more complex behaviour (see Corsini
et al. 1999; Cinzano et al. 1999).

\medskip

\noindent
{\it (ii)} Galaxies for which \Vg$\;\simeq\;$\Vs\ and
\sg$\;\simeq\;$\ss\ over an extended radial range.
This is the case of the inter\-mediate-to-late type spirals NGC 470,
NGC 949, NGC 1160, and NGC 2541, NGC 3810, and of the Sab NGC 3705.
In these disc-dominated galaxies the motions of the ionized gas and
stars are dominated by rotation as we can infer from their low
$\sigma\la50$ \kms\ and large $(V/\sigma)_{\it max}\ga2$.

\medskip

The Sab spiral NGC 3368 has intermediate properties (i.e.,
\Vg$\;>\;$\Vs\ and \sg$\;\simeq\;$\ss) between the two classes 
even though the gas rotation is quite asymmetric.  The edge-on S0 NGC
980 has a very peculiar rotation curve with \Vg$\;\ga\;$\Vs\ for
$|r|\la2''$ and \Vg$\;<\;$\Vs\ elsewhere.  These cases can be
explained if the gas disc is warped and not aligned with the plane of
the stellar disc.  Further observations on different position
angles are needed to derive a detailed modelling. For NGC 224 and NGC
5854 we have no gas kinematics to perform a comparison with stars.

\begin{figure}[h]
\centerline{\psfig{figure=MS10597f3.eps1,width=8.5cm}} 
\caption{Velocity dispersion of stars and ionized gas measured in the
centre ({\it upper panel\/}) and at $R_e/4$ ({\it lower panel\/}) of
the disc galaxies studied in this paper, Bertola et al. (1995), Fisher
(1997) and Corsini et al. (1999). The different symbols refer to the
different morphological types (as they appear in RC3).  The {\it
continuous lines\/} correspond to \sg$\;=\;$\ss. The ranges where
$\sigma\le60$ \kms\ are marked as a reference.}
\label{fig:sigma_gas_vs_stars}  
\end{figure}

In the remaining galaxies of our sample gas and stellar kinematics
suffers from the presence of kinematically decoupled components.  Two
counter-rotating stellar components have been found by Pompei \&
Terndrup (1998) in the edge-on Sb NGC 2683.  In the nuclear region of
NGC 2841 the ionized gas is rotating perpendicularly with respect to
the stars and a fraction of bulge stars are counter-rotating with
respect to the rest of the galaxy (Sil'Chenko, Vlasyuk \& Burenkov
1997).  In the centre of NGC 3031 our gaseous kinematic data
suggest the presence of a circumnuclear Keplerian disc of ionized gas
(e.g. Bertola et al. 1998) which is consistent with the gaseous disc
observed by Devereux, Ford \& Jacoby (1997) and rotating around a
supermassive black hole (Bower et al. 1996). In NGC~7331 the possible
presence of a counter-rotating bulge has been discussed by Prada et
al. (1996) and ruled out by Bottema (1999).

\begin{figure*}  
{\psfig{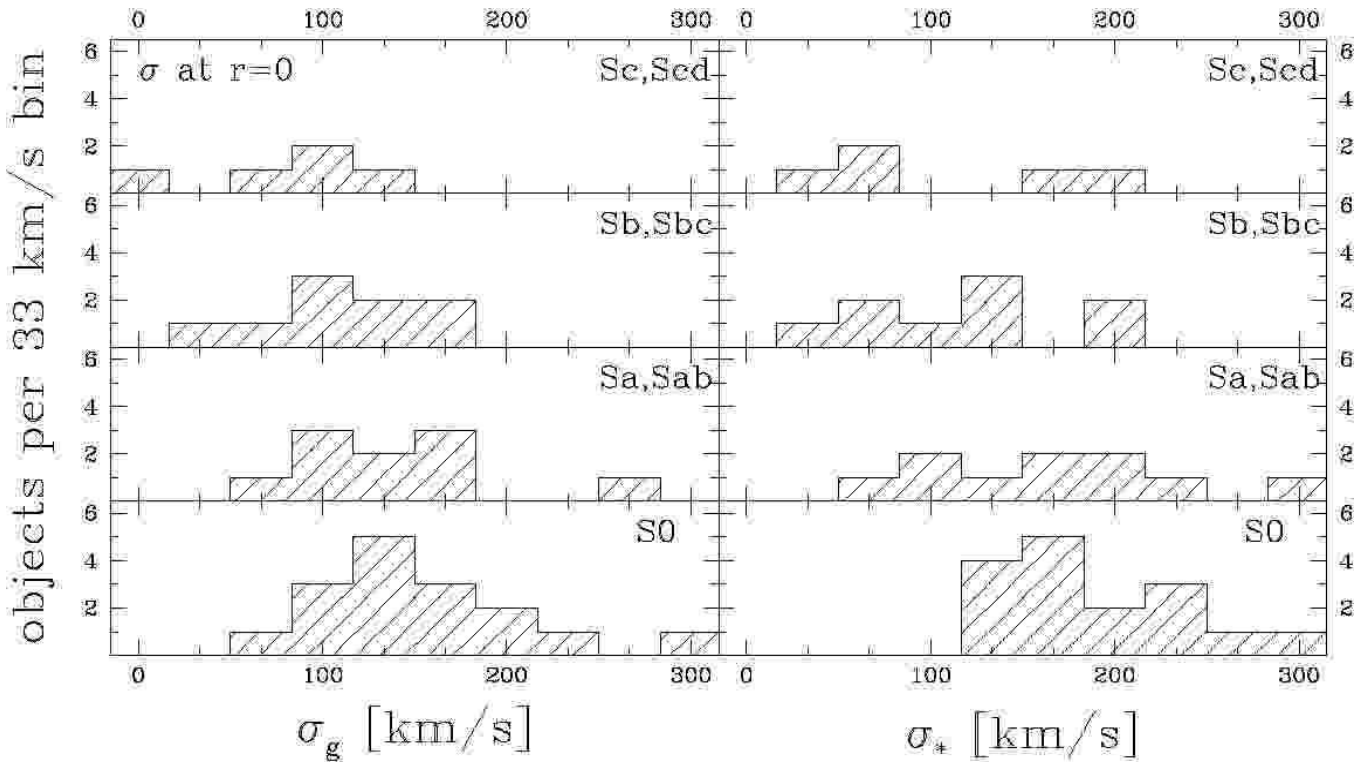}}  
\caption{Distribution of the ionized gas ({\it left panel\/}) and
  stellar velocity dispersions ({\it right panel\/}) 
  measured in the centres of the galaxies plotted in Fig. 
\protect{\ref{fig:sigma_gas_vs_stars}}. Galaxies have been sorted
  according to their RC3 morphological type. The velocity-dispersion bins
  are 33 \kms\ wide.}
\label{fig:histogram_sigma_zero}  
\end{figure*}  

\begin{figure*}  
{\psfig{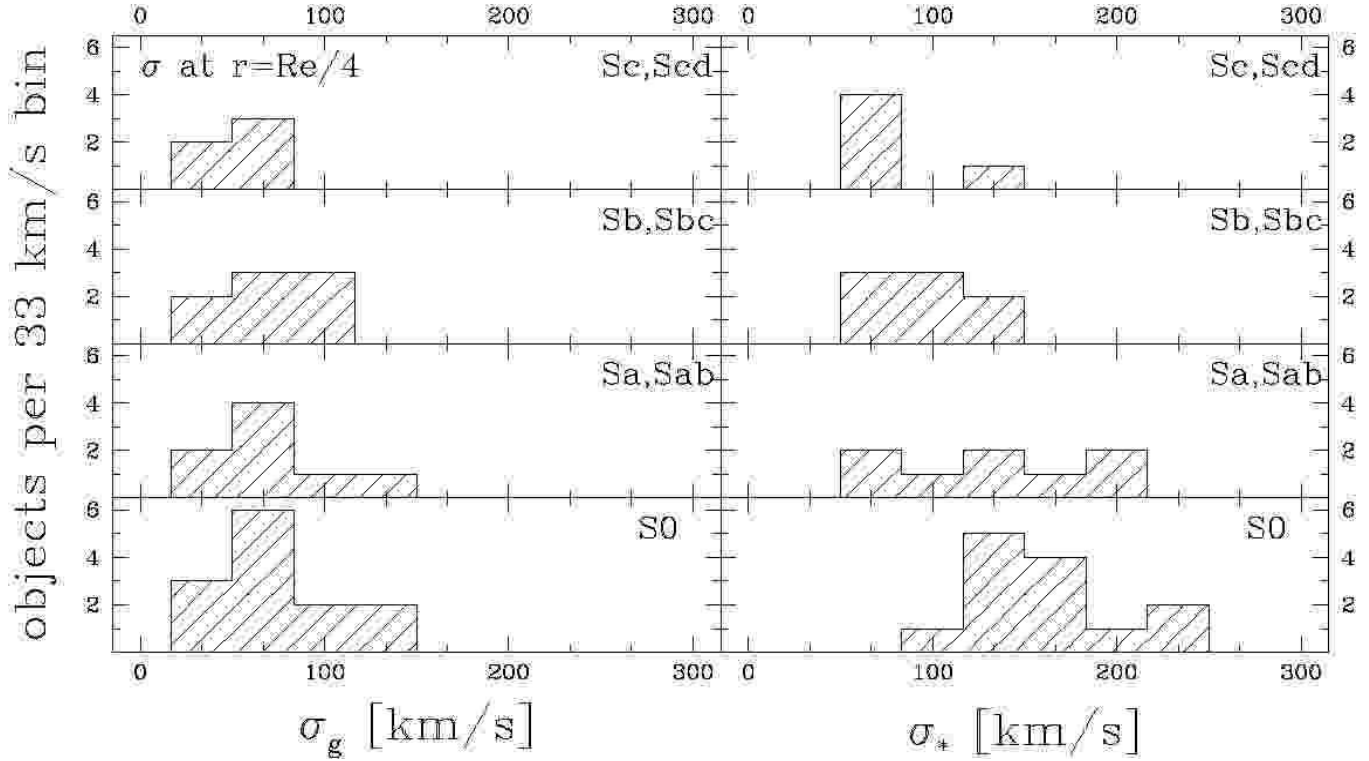}}  
\caption{As in Fig.\protect{\ref{fig:histogram_sigma_zero}} but for 
  ionized gas and stellar velocity dispersions measured at $R_e/4$.}
\label{fig:histogram_sigma_re/4}  
\end{figure*}

The recent results implying a tight $M_\bullet-\sigma$ relation of its
host spheroids (Ferrarese \& Merritt 2000) have made us look for
possible relationships between the velocity dispersion of gas and
stars.
With this aim we compiled data from a sample of about 40 disc galaxies
for which the major-axis velocity curve and velocity dispersion
profiles of both ionized gas and stars are available, by adding the
early-type disc galaxies of Bertola et al. (1995), Fisher (1997), and
Corsini et al. (1999) to the spirals of our present sample. Although
stellar and/or ionized gas kinematics have been studied in a larger
number of S0's and spirals (e.g. H\'eraudeau \& Simien 1998;
H\'eraudeau et al. 1999) we selected these few authors since only they
provide the radial trend of the gas velocity dispersion.
For each object we derived the values of \ss\ and \sg\ in the centre
and at $R_e/4$, where $R_e$ is the half-surface brightness radius of
the galaxy listed in the RC3 (Fig. \ref{fig:sigma_gas_vs_stars}).

The central values of \ss\ and \sg\ seem to be correlated, since
galaxies with higher \ss\ tend to show also higher values of \sg .
Moreover \ss\ and \sg\ cover the same range of values and can reach
values higher than $300$ \kms\ with \sg$\;\la\;$\ss . However there is
no clear dependence on the morphological type as seen in
Fig. \ref{fig:histogram_sigma_zero}.  The high central values of \sg\
may be partially due to the smearing effect of the seeing, which are
more noticeable on the gas kinematics of the early-type disc galaxies
on account of their large central velocity gradients (e.g. Rubin et
al. 1985).  On the other hand, a high central \sg\ could be also due to
intrinsic properties of the galaxy. This is the case of the broad emission
lines which are the signature of an unresolved Keplerian velocity
field due to a gaseous disc rotating around a supermassive black hole
(e.g. Bertola et. 1998; Maciejewski \& Binney 2000).

At $R_e/4$ we find \sg$\la100$ \kms\ in almost all the objects, while
\ss\ ranges between $40$ and $240$ \kms . This correlates with the
galaxy type as seen in Fig. \ref{fig:histogram_sigma_re/4}.
The low value of \sg\ indicates that we are observing dynamically cold
gas, which is rotating in the disc component. In addition, there are a
few S0's and early-type spirals in which \sg\ is much higher
(\sg$\;\simeq120$ \kms) than expected from thermal motions or
small-scale turbulence.  Such a high \sg\ can not be explained as
the result of seeing smearing of velocity gradients since it is
measured at a distance $R_e/4$ which is larger than 5 seeing FWHM's
for all the sample objects. We suggest that the high-\sg\ galaxies are
good candidates to host dynamically hot ionized gas as in the case of
the S0 NGC 4036 (Bertola et al. 1995; Cinzano et al. 1999),  even
if the question whether pressure-supported gas is related to the
dynamics of the bulge stars is still open (Pignatelli et al. 2001).
The Hubble Type - \ss\ relation observed at $R_e/4$ is an indication
that at this radius the stellar kinematics of early and late-type disc
galaxies dominated by bulge and disc component, respectively. In
late-type spirals, which host low or negligible bulges,
\ss$\;\simeq\;$\sg$\;\simeq\;$50 \kms.

\acknowledgements

This research was partially based on data from ING archive, and has
made use of the Lyon-Meudon Extragalactic Database (LEDA) and of the
NASA/IPAC Extragalactic Database (NED).
We are grateful to Prof. P.A. Strittmatter, Director of the Steward
Observatory and to Dr. G.V. Coyne, S.J., Director of the Vatican
Observatory for the allocation of time for our observations. The
Multiple Mirror Telescope is a joint facility of the Smithsonian
Institution and the University of Arizona.  The Vatican Advanced
Technology Telescope is the Alice P. Lennon Telescope and the Thomas
J. Bannan Astrophysics Facility.
WWZ acknowledges support of the {\sl Jubil\"aum\-sfonds der
Oesterreichischen Nationalbank} (grant 7914).  JEB acknowledges
support of the Spanish DGES (grant PB97-0214).
JEB, JCVB and WWZ acknowledges the support of this research project in the framework of the 
 Austrian-Spanish {\it Acci\'on Integrada} (project No. 20/2000).

\appendix
\section{Comparison with literature} 
\label{sec:comparison}

\begin{figure*}
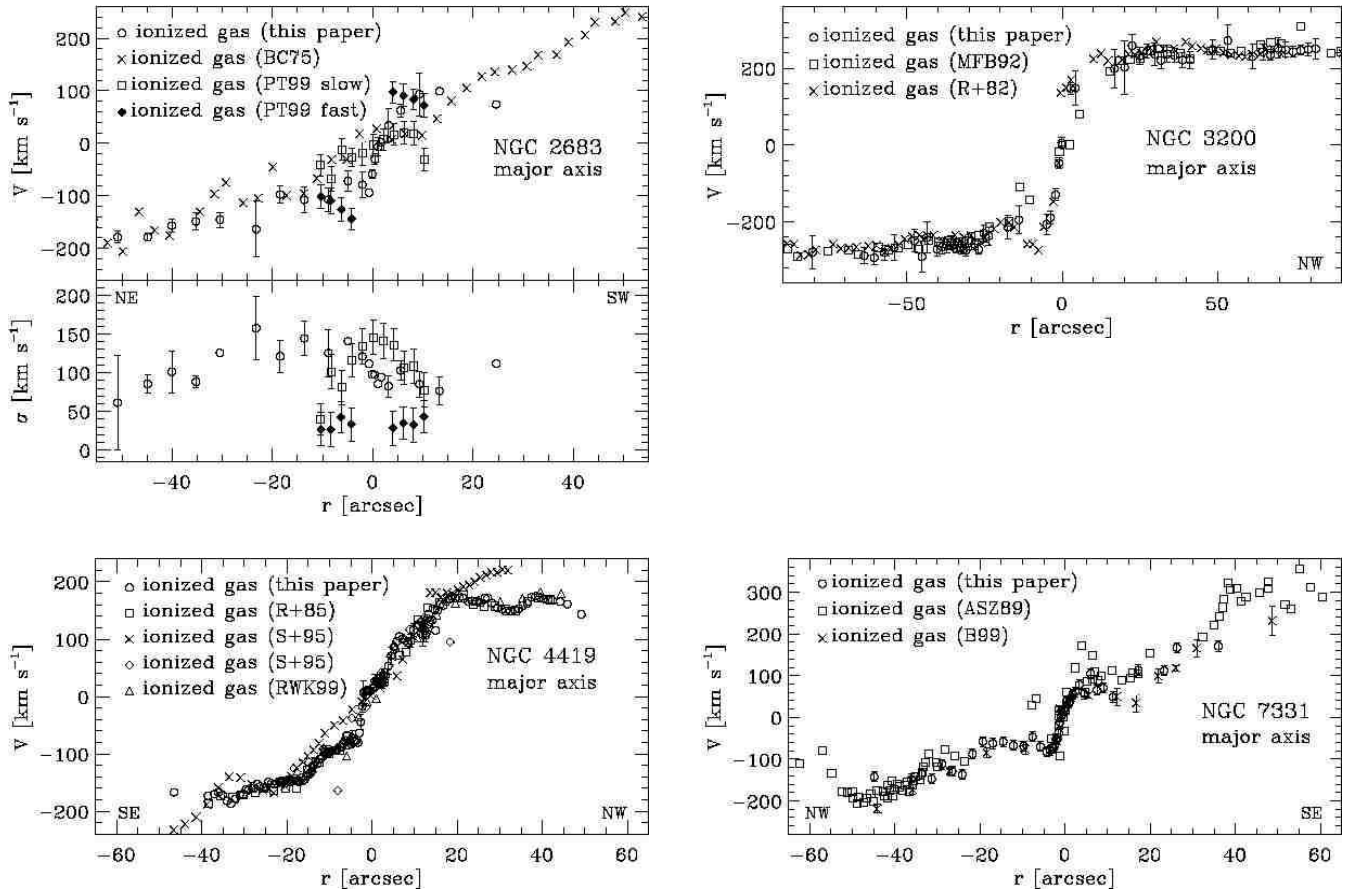

\begin{minipage}[t]{8.5cm}
\vspace{0pt}
\psfig{figure=MS10597fA1a.eps1,width=8.5cm} 
\end{minipage}
\hspace*{0.5cm}
\begin{minipage}[t]{8.5cm}
\vspace{0pt}
\psfig{figure=MS10597fA1b.eps1,width=8.5cm}
\end{minipage}

\vspace*{0.5cm}
\begin{minipage}[t]{8.5cm}
\vspace{0pt}
\psfig{figure=MS10597fA1c.eps1,width=8.5cm}
\end{minipage}
\hspace*{0.5cm}
\begin{minipage}[t]{8.5cm}
\vspace{0pt}
\psfig{figure=MS10597fA1d.eps1,width=8.5cm}
\end{minipage}
\caption{The ionized gas velocities derived in this study for NGC
  2683, NGC 3200, NGC 4419 and NGC 7331 compared with those obtained
  by other authors:  ASZ89 = Afanasiev et al. 1989; B99 = Bottema 1999; 
  BC75 = Barbon \& Capaccioli 1975; MFB92 = Mathewson et al. 1992; 
  R+82 = Rubin et al. 1982; R+85 = Rubin et al. 1985; 
  RWK99 = Rubin et al. 1999; S+95 = Sperandio et al. 1995.}
\label{fig:gascomparison}
\end{figure*} 

\begin{figure*}
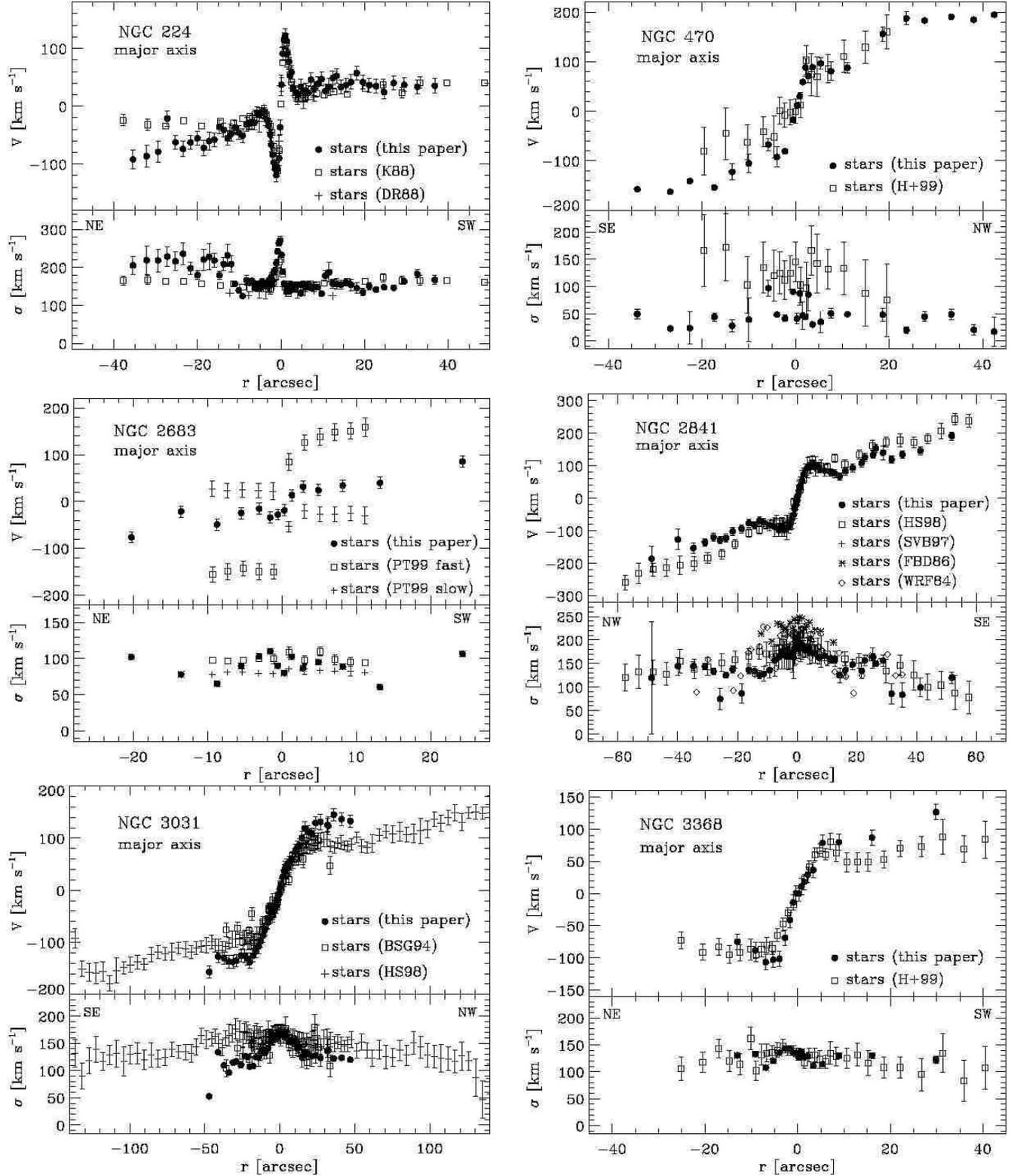

\hbox{
\psfig{figure=MS10597fA2a.eps1,width=8.5cm}\hspace*{0.5cm}
\psfig{figure=MS10597fA2b.eps1,width=8.5cm}}
\hbox{
\psfig{figure=MS10597fA2c.eps1,width=8.5cm}\hspace*{0.5cm}
\psfig{figure=MS10597fA2d.eps1,width=8.5cm}}
\hbox{
\psfig{figure=MS10597fA2e.eps1,width=8.5cm}\hspace*{0.5cm}
\psfig{figure=MS10597fA2f.eps1,width=8.5cm}}
\caption{The stellar velocities and velocity dispersions derived in
  this study for NGC 224, NGC 470, NGC 2841, NGC 3031, NGC 3368, NGC
  3810, NGC 5854 and NGC 7331 compared with those obtained by other
  authors: B99 = Bottema 1999; B+93 = Bower et al.; 1993 BSG94 = Bender et
  al. 1994; DR88 = Dressler \& Richstone 1988; FBD86 = Fillmore et
  al. 1986; H+99 = H\'eraudeau et al. 1999; H+00 = Haynes et al. 2000;
  HS98 = H\'eraudeau \& Simien 1998; K88 = Kormendy 1988; P+96 = Prada et
  al. 1996; SP97 = Simien \& Prugniel 1997; SVB86 = Sil'Chenko et
  al. 1997; WRF84 = Whitmore et al. 1984.}
\label{fig:starcomparison_a}
\end{figure*} 

\setcounter{figure}{1}
\begin{figure*}
\begin{minipage}[t]{8.5cm}
\vspace{0pt}
\psfig{figure=MS10597fA2g.eps1,width=8.5cm}
\end{minipage}
\hspace*{0.5cm}
\begin{minipage}[t]{8.5cm}
\vspace{0pt}
\psfig{figure=MS10597fA2h.eps1,width=8.5cm}\hspace*{0.5cm}
\end{minipage}
\vspace*{0.5cm}
\begin{minipage}[t]{8.5cm}
\vspace{0pt}
\psfig{figure=MS10597fA2i.eps1,width=8.5cm}
\end{minipage}
\hspace*{0.5cm}
\begin{minipage}[t]{8.5cm}
\vspace{0pt}
\psfig{figure=MS10597fA2j.eps1,width=8.5cm}
\end{minipage}
\caption{(continue)}
\label{fig:starcomparison_b}
\end{figure*}

In this section we perform a comparison between the gaseous and
stellar kinematical data
we obtained for the sample galaxies with
the velocity curves and velocity dispersion profiles available in
literature in order to assess the accuracy and reliability of our measurements.
In most cases differences between different authors are due to slit
centring and positioning and to the different analysis 
techniques, or both (see Fisher 1997 for a discussion).   

\medskip

As far as our sample galaxies are concerned, ionized gas velocity
curves have been already measured along the major axis of NGC 2683,
NGC 3200, NGC 3898, NGC 4419, and NGC 7331. Apart from NGC 3898
presented in Pignatelli et al. (2001) the other cases are discussed
here briefly and shown in Fig. \ref{fig:gascomparison}.

\noindent
{\bf NGC 2683:} The complex kinematics of this galaxy has been
unveiled by Pompei \& Terndrup (1999) who isolated two kinematically
distinct gaseous components giving rise to a `figure-of-eight'
velocity curve. The fast and the slow-rotating components are
unresolved in our spectrum as well as in that of Barbon \& Capaccioli
(1975). We therefore measured intermediate \Vg 's and higher
\sg 's.

\noindent
{\bf NGC 3200:} The agreement between our \Vg\ and that measured by
Rubin et al. (1982) is excellent. This is also true for Mathewson et
al. (1992) in the outer regions.
The shallower central gradient can be explained taking into
account for their lower spatial resolution.

\noindent
{\bf NGC 4419:} Our data closely matches those obtained by previous
authors. This is not the case for only of the \Vg\ rotation curves
given by Sperandio et al. (1995). In this case their lower spatial
resolution produces the observed shallower central gradient but it can
not account for the strong discrepancy we see at large radii on both
sides.

\noindent
{\bf NGC 7331:} Our \Vg 's matches those by Bottema (1999). A
difference in the heliocentric systemic velocity and the different
position angle of the slit (P.A.$=170\dg$) may explain the shift the
\Vg\ curve by Afanas'ev, Sil'Chenko \& Zasov (1989).

\medskip

Major-axis stellar kinematics have been previously published for NGC 224, NGC
470, NGC 772, NGC 2683, NGC 2841, NGC 3031, NGC 3368, NGC 3810, 
NGC 3898, NGC 5854 and NGC 7331. They are compared with our data in
Fig. \ref{fig:starcomparison_a} except for the cases of NGC 772 and NGC
3898 which we analyzed in Pignatelli et al. (2001).

\noindent
{\bf NGC 224:} The discrepancy observed along the NE axis between our
data and those by Kormendy (1988) and by Dressler and Richstone (1988)
may be the result of an incorrect sky subtraction in our data.
An overestimation of the sky level due to the large size of the galaxy
covering all the slit area may produce the higher \ss\ and shift in \Vs\ we
actually measure.

\noindent
{\bf NGC 470:} The \Vs 's we measured are consistent with those of
H\'eraudeau et al. (1999) while less satisfactory is the comparison
between our and their \ss .  In particular their values range between
about $70$ and $170$ \kms , whereas we measured \ss$\;\la50$ \kms\ at
almost all radii (probably due to a template mismatching effect).

\noindent
{\bf NGC 2683:} We do not resolve the counter-rotating stellar 
components observed by Pompei \& Terndrup (1999) because we have no enough
spectral resolution and either a good S/N in our spectra.  

\noindent
{\bf NGC 2841:} 
In the centre the \Vs\ value we obtained is within
the scatter of the other data sets, the same is true for \ss . however further
out from the nucleus our \Vs\ and \ss\ are are somewhat lower
than those found in literature.

\noindent
{\bf NGC 3031:} We measure the same \Vs\ gradient as Bender, Saglia \&
Gerhard (1994) and H\'eraudeau \& Simien (1998) in the inner
$|r|\la10''$. Further out our \Vs\ continues to increase. The
differences between the three  \Vs\ sets are as large as 
$50$--$80$ \kms .  In the same radial region
our \ss\ agrees with the velocity dispersions by Bender et al. (1994)
but are about $50$ \kms\ lower than those by H\'eraudeau \& Simien
(1998). \ss\ measurements do coincide in the centre. These differences
in \Vs\ and \ss\ are due to the different instrumental setup used for the
different observations. We used a very spatial resolution, so the velocity
gradient we measure is more accurate than that of the other authors. 
The differences  in  the values of \ss\  between our measurements and 
those obtained by H\'eraudeau \& Simien (1998) 
are probably due  to a template mismatching effect
 
\noindent
{\bf NGC 3368, NGC 3810 and NGC 5854:} Our major-axis kinematics are
in good agreement with the literature.  This is also true for the \Vs\
we measured along the minor axis of NGC 5854.

\noindent
{\bf NGC 7331:} The \Vs\ gradient measured by H\'eraudeau \& Si\-mien
(1998) is shallower than that by us and other authors. We suggest it
may be due to a different position of the slit.  Our \ss\ radial
profile shows a faster decrease and at larger radii it is marginally
consistent with that by Bower et al. (1993).

\medskip

For some of the sample galaxies velocity fields for the cool gaseous
component have been obtained using CO molecular lines and/or \hi\
21-cm line and can be compared to our ionized-gas velocity curves to
have some insights into the inner-to-outer gas distribution and
motion.

\noindent
{\bf NGC 2541:} The \hi\ data by Broeils \& van Woerden (1994) show a
symmetric outer rotation curve with no particular peculiarities.

\noindent
{\bf NGC 2683:} The \hi\ position-velocity diagram by Broeils \& van
Woerden (1994) shows two kinematically distinct components giving to
the diagram a `figure-of-eight' appearance. These components may be
associated to the fast and slow-rotating components observed by Pompei
\& Terndrup (1998) both in the stellar and ionized-gas velocity
curves. Figure-of-eight velocity curve have been explained by Kuijken
\& Merrifield (1995) as due to the presence of a bar (see other
examples in Vega Beltr\'an et al. 1997; Bureau \& Freeman 1999).

\noindent
{\bf NGC 2841:} Sofue et al. (1999) obtained an extended and
well-sampled inner-to-outer rotation curve by combining \ha, CO and
\hi\ observations. Rotation attains a sharp maximum near the centre
and flattens outwards and asymmetries seem to be confined in the
radial region we observed.

\noindent
{\bf NGC 3031:} Sofue (1997) combined different optical and radio data
sets to trace gas rotation of this spiral galaxy out to more than
$20'$. There is no evidence of kinematical decoupling between gas and
stars even if our higher-resolution data show that in the inner $\pm1'$
the gas velocity curve is highly disturbed and less regular than the
stellar one.

\noindent
{\bf NGC 3368:} The \hi\ and CO velocity fields have been derived by
Schneider (1989) and Sakamoto et al. (1999) respectively.  The large
fraction of \hi\ is distributed outside the optical disk indicating
the possible capture of intergalactic gas.  On the contrary CO is
concentrated towards the inner regions and its asymmetric
position-velocity diagram matches our \oiii\ velocity curve. Gas
infall due to the galaxy bar and interactions has been considered by
Sakamoto et al. (1999) to explain CO distribution.

\noindent
{\bf NGC 3898:} The \hi\ distribution and velocity field have been
studied in detail by van Driel \& van Woerden (1994).  The comparison
of these data with our ionized-gas kinematics and the
\ha -imaging by Pignatelli et al. (2001) suggests that
both ionized and neutral hydrogen have a regular velocity field and a
smooth distribution.

\noindent
{\bf NGC 7331:} Sofue (1997) obtained a rotation curve for the gaseous
component from CO and \hi\ lines. The global rotation appears normal
with no peculiar behaviour. The is no gas component associated to the
counter-rotating bulge claimed by Prada et al. (1996).

\clearpage

\begin{landscape}
\begin{table}
\caption[]{Kinematic properties of the sample galaxies}  
\begin{flushleft}  
\begin{scriptsize}
\begin{tabular}{lrrcrrcccccccrrcll}  
\hline  
\noalign{\smallskip}   
\multicolumn{1}{c}{Object} &   
\multicolumn{2}{c}{$\sigma(0)         $}&  
\multicolumn{1}{c}{}                   &   
\multicolumn{2}{c}{$\sigma(R_e/4)  $} &
\multicolumn{1}{c}{}                   &      
\multicolumn{2}{c}{$(\Delta V/\Delta R)(0)       $}&  
\multicolumn{1}{c}{}                   &    
\multicolumn{2}{c}{$(\Delta V/\Delta R)(R_e/4)$}&   
\multicolumn{1}{c}{}                   &    
\multicolumn{2}{c}{$V(R_e/4)       $} &
\multicolumn{1}{c}{} &
\multicolumn{2}{c}{$R_{\it last}$}  \\  
\noalign{\smallskip}
\cline{2-3} \cline{5-6} \cline{8-9} \cline{11-12} \cline{14-15} \cline{17-18}
\noalign{\smallskip}   
\multicolumn{1}{c}{}      &   
\multicolumn{1}{c}{gas}   &  
\multicolumn{1}{c}{stars} &  
\multicolumn{1}{c}{}      &    
\multicolumn{1}{c}{gas}   &   
\multicolumn{1}{c}{stars} &  
\multicolumn{1}{c}{}      &
\multicolumn{1}{c}{gas}   &      
\multicolumn{1}{c}{stars} &   
\multicolumn{1}{c}{}      &    
\multicolumn{1}{c}{gas}   &  
\multicolumn{1}{c}{stars} &
\multicolumn{1}{c}{}      &   
\multicolumn{1}{c}{gas}   &  
\multicolumn{1}{c}{stars} &
\multicolumn{1}{c}{}      &
\multicolumn{1}{c}{gas}   &  
\multicolumn{1}{c}{stars} \\    
\noalign{\smallskip}   
\multicolumn{1}{c}{[name]}  &   
\multicolumn{2}{c}{[\kms]}  &  
\multicolumn{1}{c}{}        &    
\multicolumn{2}{c}{[\kms]}  &   
\multicolumn{1}{c}{}  &   
\multicolumn{2}{c}{[\kmspc]}&      
\multicolumn{1}{c}{}        &    
\multicolumn{2}{c}{[\kmspc]}&
\multicolumn{1}{c}{}        &      
\multicolumn{2}{c}{[\kms]}&
\multicolumn{1}{c}{}      &
\multicolumn{2}{c}{[$R/R_{25}$]}     \\  
\noalign{\smallskip}   
\multicolumn{1}{c}{(1)}   &   
\multicolumn{1}{c}{(2)}   &  
\multicolumn{1}{c}{(3)}   &  
\multicolumn{1}{c}{}      &    
\multicolumn{1}{c}{(4)}   &   
\multicolumn{1}{c}{(5)}   &  
\multicolumn{1}{c}{}      &   
\multicolumn{1}{c}{(6)}   &      
\multicolumn{1}{c}{(7)}   &   
\multicolumn{1}{c}{}      &    
\multicolumn{1}{c}{(8)}   &  
\multicolumn{1}{c}{(9)}  &
\multicolumn{1}{c}{}      &
\multicolumn{1}{c}{(10)}  &
\multicolumn{1}{c}{(11)}  &
\multicolumn{1}{c}{}    &
\multicolumn{1}{c}{(12)}  &
\multicolumn{1}{c}{(13)}  \\     
\noalign{\smallskip}   
\hline  
\noalign{\smallskip}   
NGC~224 &\multicolumn{1}{c}{\ ---}&$244\pm 25$&&\multicolumn{1}{c}{\ ---}&\multicolumn{1}{c}{\ ---}& \multicolumn{1}{c}{---} && $63.29\pm 0.720$& &\multicolumn{1}{c}{---}&\multicolumn{1}{c}{---} &&\multicolumn{1}{c}{---} & \multicolumn{1}{c}{---} && --- & 0.006 \\ 
NGC~470 &$ 84\pm\ \, 8$&$ 56\pm    31$ &&$ 37\pm    12$ &$ 63\pm    18$ && $0.563\pm 0.024$ & $0.275\pm 0.010$ && $0.095\pm 0.010$ & $0.036\pm 0.014$ && $122\pm 13$ & $104\pm 15$ && 0.6 & 0.5\\
NGC~772 &$128\pm    13$&$124\pm\ \, 5$ &&$ 69\pm    12$ &$ 94\pm    55$ && $0.283\pm 0.022$ & $0.195\pm 0.012$ && $0.073\pm 0.009$ & $0.051\pm 0.010$ && $239\pm 22$ & $170\pm 28$ &&0.2 & 0.1  \\
NGC~949 &$ 37\pm    19$&$ 32\pm\ \, 5$ &&\multicolumn{1}{c}{\ ---}&\multicolumn{1}{c}{\ ---}&& $0.115\pm 0.060$ & $0.078\pm 0.011$ && \multicolumn{1}{c}{---}&\multicolumn{1}{c}{---}&&\multicolumn{1}{c}{---} &\multicolumn{1}{c}{---} && 0.6 & 0.6  \\
NGC~980 &$212\pm\ \, 7$&$226\pm    12$ &&\multicolumn{1}{c}{\ ---}  &\multicolumn{1}{c}{\ ---}&& $0.222\pm 0.020$ & $0.197\pm 0.004$ && \multicolumn{1}{c}{---}&\multicolumn{1}{c}{---} &&\multicolumn{1}{c}{---} &\multicolumn{1}{c}{---} && 0.8 & 0.8  \\
NGC~1160&$109\pm    25$&$ 24\pm    10$ &&$ 50\pm    29$ &$ 51\pm    30$ && $0.053\pm 0.027$ & $0.042\pm 0.021$ && $0.031\pm 0.017$ & $0.017\pm 0.006$ && $ 75\pm 19$ & $ 48\pm 13$ && 0.8 & 0.6\\
NGC~2541&$ 16\pm    16$&$ 53\pm    10$ &&$ 29\pm    19$ &$ 59\pm    10$ && $0.092\pm 0.071$ & $0.066\pm 0.066$ && $0.027\pm 0.018$ & $0.067\pm 0.038$ && $ 45\pm 10$ & $ 43\pm 15$ && 0.2 & 0.08\\
NGC~2683&$ 97\pm\ \, 5$&$ 83\pm\ \, 7$ &&$110\pm    27$ &$ 71\pm    33$ && $0.998\pm 0.304$ & $0.527\pm 0.069$ && $0.294\pm 0.048$ & $0.096\pm 0.135$ && $105\pm 16$ & $ 35\pm 47$ && 0.2 & 0.1\\
NGC~2841&$141\pm    11$&$197\pm    18$ &&$104\pm    11$ &$118\pm    29$ && $0.697\pm 0.326$ & $0.934\pm 0.020$ && $0.150\pm 0.044$ & $0.120\pm 0.019$ && $160\pm 20$ & $130\pm 34$ && 0.3 & 0.2\\
NGC~3031&$251\pm\ \, 5$&$173\pm\ \, 5$ &&\multicolumn{1}{c}{\ ---}&\multicolumn{1}{c}{\ ---}&&$7.452\pm 0.174$ & $1.762\pm 0.160$ &&\multicolumn{1}{c}{---}&\multicolumn{1}{c}{---}&& \multicolumn{1}{c}{---}& \multicolumn{1}{c}{---}&& 0.05 & 0.06\\
NGC~3200&$ 82\pm    21$&$197\pm    35$ &&$ 38\pm    19$ &$132\pm    53$ && $0.267\pm 0.021$ & $0.095\pm 0.019$ && $0.103\pm 0.011$ & $0.068\pm 0.039$ && $193\pm 19$ & $128\pm 60$ && 0.6 & 0.4\\
NGC~3368&$140\pm\ \, 5$&$129\pm\ \, 5$ &&$117\pm    32$ &$128\pm    11$ && $0.974\pm 0.252$ & $0.566\pm 0.074$ && $0.249\pm 0.030$ & $0.119\pm 0.051$ && $216\pm 19$ & $103\pm 30$ && 0.09 & 0.1\\
NGC~3705&$122\pm\ \, 5$&$109\pm\ \, 9$ &&$ 58\pm    32$ &$ 88\pm\ \, 6$ && $0.681\pm 0.120$ & $0.327\pm 0.072$ && $0.260\pm 0.115$ & $0.116\pm 0.026$ && $113\pm 37$ & $ 50\pm 10$ && 0.08& 0.07\\
NGC~3810&$110\pm    15$&$ 58\pm    12$ &&$ 54\pm    25$ &$ 70\pm\ \, 5$ && $0.255\pm 0.272$ & $0.171\pm 0.070$ && $0.147\pm 0.085$ & $0.151\pm 0.089$ && \multicolumn{1}{c}{---}& $79\pm 25$ && 0.08 & 0.06\\
NGC~3898&$103\pm    12$&$223\pm    15$ &&$ 68\pm\ \, 5$ &$161\pm    16$ && $0.338\pm 0.044$ & $0.336\pm 0.030$ && $0.328\pm 0.026$ & $0.226\pm 0.023$ && $199\pm 10$ & $140\pm 12$ && 0.7 & 0.2\\
NGC~4419&$104\pm\ \, 6$&$ 98\pm    12$ &&$ 35\pm\ \, 5$ &$ 76\pm    24$ && $0.152\pm 0.014$ & $0.123\pm 0.040$ && $0.180\pm 0.011$ & $0.125\pm 0.022$ && $ 90\pm 11$ & $ 63\pm 16$ && 0.5 & 0.5\\
NGC~5064&$114\pm\ \, 8$&$202\pm    11$ &&$ 54\pm\ \, 6$ &$210\pm    21$ && $0.366\pm 0.015$ & $0.240\pm 0.018$ && $0.189\pm 0.013$ & $0.100\pm 0.018$ && $225\pm 15$ & $132\pm 20$ && 0.1 & 0.6\\
NGC~5854& \multicolumn{1}{c}{\ ---} & $109\pm\ \, 7$ &  &\multicolumn{1}{c}{\ ---}&$ 99\pm    16$ &&  \multicolumn{1}{c}{---} & $0.100\pm 0.030$ && \multicolumn{1}{c}{---} &   $0.100\pm 0.075$ && \multicolumn{1}{c}{---} & $ 38\pm 25$ && --- & 0.5 \\
NGC~7331&$ 90\pm    11$&$141\pm\ \, 5$ &&$ 88\pm    13$ &$ 85\pm\ \, 5$ && $0.307\pm 0.034$ & $0.427\pm 0.022$ && $0.079\pm 0.023$ & $0.185\pm 0.011$ && $ 79\pm 20$ & $186\pm 10$ && 0.1 & 0.3\\
NGC~7782&$169\pm    31$&$193\pm\ \, 9$ &&$ 55\pm    25$ &$120\pm    15$ && $0.539\pm 0.015$ & $0.228\pm 0.018$ && $0.114\pm 0.005$ & $0.084\pm 0.016$ && $266\pm 10$ & $194\pm 33$ && 0.6 & 0.6\\
\noalign{\smallskip}  
\hline  
\noalign{\smallskip}  
\noalign{\smallskip}  
\noalign{\smallskip}  
\end{tabular} 
\end{scriptsize}
\begin{footnotesize}  
\begin{minipage}{23cm}  
 NOTES -- 
Cols.(2-3): central velocity dispersion of ionized gas and stars.
Cols.(4-5): velocity dispersion of ionized gas and stars at $R_e/4$.
Cols.(6-7): central velocity gradient of ionized gas and stars.
Cols.(8-9): velocity gradient of ionized gas and stars at $R_e/4$. 
Cols.(10-11): ionized-gas and stellar rotation velocity at $R_e/4$ obtained 
              from the observed velocity corrected for systemic
              velocity and inclination given in Tab. 1.
Cols.(12-13): Extent of the ionized-gas and stellar kinematic radial profiles
              obtained in this paper in units of $R_{25}$. 
\end{minipage}  
\end{footnotesize}    
\label{tab:results}  
\end{flushleft}  
\end{table}  
\end{landscape}

\end{document}